# Deep Reinforcement Learning for Optimal Control of Space Heating

Adam Nagy[1], Hussain Kazmi[1,2], Farah Cheaib[3], Johan Driesen[2]
[1] KU Leuven, Belgium
[2] Enervalis, Houthalen-Helchteren, Belgium
[3] Estabanell Energia, Spain

## Abstract
Classical methods to control heating systems are often marred by suboptimal performance, inability to adapt to dynamic conditions and unreasonable assumptions e.g. existence of building models. This paper presents a novel deep reinforcement learning algorithm which can control space heating in buildings in a computationally efficient manner, and benchmarks it against other known techniques. The proposed algorithm outperforms rule based control by between 5-10% in a simulation environment for a number of price signals. We conclude that, while not optimal, the proposed algorithm offers additional practical advantages such as faster computation times and increased robustness to non-stationarities in building dynamics.

## Introduction

Anthropogenic climate change is amongst the key challenges facing humanity in the 21st century (Intergovernmental Panel on Climate Change, 2015). The use of fossil fuels to meet energy demand is a key contributing factor to the changing climate (Höök, 2013). Energy consumed to condition living and working spaces is a large contributor to this; being responsible for over one third of total greenhouse gas (GHG) emissions in the EU (Filippidou, 2017). Rapidly increasing levels of population, urbanization and electrification mean this situation might further worsen unless corrective action is taken.

Recent pieces of legislation such as the Energy Performance of Buildings Directive, EPBD (and its recast) have mapped the regulatory pathway towards higher efficiency buildings in Europe (Hamdy, 2013). In particular, mandating all new buildings to consume 'nearly zero energy' has led to market innovations in both building façade and heating equipment. Improved insulation has meant less energy being lost to the ambient in the form of transmission or infiltration losses. Heat exchangers coupled with ventilation units have further reduced ventilation losses while the use of higher efficiency heat pumps or district heating networks has reduced the primary energy required to condition living spaces.

In this discourse, the importance of occupants on end energy demand has emerged as a critical aspect (Oldewurtel F. D., 2013). With buildings consuming much less energy to provide the same service as before, a rebound effect of sorts has been observed in many cases. This human performance gap is evidenced as people raising their expectations of thermal comfort. It also manifests in the consistently lower actual energy performance of buildings when compared to their theoretical rating and is especially problematic in modern, higher efficiency buildings (Majcen, 2015).

One possibility to reduce energy consumption for space conditioning is automatically controlling the heating equipment while respecting occupant defined comfort bounds (Gill, 2014). These automatic control mechanisms have to go beyond classical rule based control (RBC) formulations, which are usually implemented as naïve hysteresis loops, reheating the building every time a temperature threshold is met.

Model Predictive Control (MPC) offers an obvious improvement over RBC in the quality of control by bringing anticipative prowess to the control procedure (Afram, 2014). The building is reheated in this case while optimizing towards a secondary objective such as reducing costs. MPC, while outperforming RBC, introduces complexity to the system and assumes the existence of a model explaining system dynamics. This model is often based on building physics models and has to be constructed offline – an assumption that is usually not valid for residential buildings as creating a detailed and accurate model would be too costly in practice. Furthermore, BIM and energy models, even when they are available, can have significant discrepancies based on theoretical and practical performance (Majcen, 2015). Data-driven control can alleviate the limitations caused by requiring an accurate model (Kazmi H. e., 2016); however, the computational expense to continuously learn and plan can be significant.

In recent years, reinforcement learning (RL) has emerged as a viable alternative to MPC in many domains. The allure of RL lies in its ability to approach the level of savings offered by model predictive controllers while learning directly from sensor data, i.e. not requiring the presence of a model beforehand. A number of reinforcement learning algorithms have been proposed in literature, which can be broadly classified as model-based and model-free RL algorithms (Sutton, 2017), (Kazmi H. F., 2017). While both are data driven and have their advantages and disadvantages, in this paper we focus on developing a model-free controller because of the computational advantages this class of algorithms offers. We then compare the performance of this model-free controller with an ideal MPC (which has access to the true system dynamics model) and a data driven model-based controller, which can be classified as either a model-based RL algorithm or data-driven MPC.

To the best of our knowledge, there has been no thorough comparative study on the pros and cons of different reinforcement based controllers. Some studies proposing reinforcement learning strategies for different aspects of building control have however recently appeared (Barrett, 2015), (Wei, 2017), (Ali, 2017) (Kazmi, 2018). Without knowing how these algorithms stack up against each other or known model-based controllers, there is no way for practitioners to adopt one or the other. We intend this to address this situation by quantifying the performance of state of the art RL algorithms for optimal control of buildings. In doing so, it also hopes to provide future directions for research in data driven building control and help researchers apply and report their findings in more standardized settings.

A note on terminology is necessary here. While the distinction between MPC and model-free RL is obvious, the difference between model-based RL and data-driven MPC is harder to define. There are, in our opinion, two dimensions to this. The first is exploration and the second is policy-side learning. Active exploration to improve the learnt system dynamics model is usually exclusively an RL construct and is seldom seen in data-driven MPC implementations. While exploration might sometimes lead to degraded performance momentarily, it improves performance over the long run. Likewise, while both data-driven MPC and model-based RL algorithms improve their representation of the system dynamics through observation data, model-based RL algorithms can go an extra step and learn the optimal policy as a corollary to this learning. This is evidenced in the popular family of Dyna algorithms (Sutton, 2017) and can help improve the quality and speed of control.

The rest of the paper is organized as follows: in the next section, we explain the simulation framework used to generate building dynamics and the methodology followed by the model-free and model-based reinforcement learning algorithms. We then discuss results obtained along the two identified dimensions of interest: energy efficiency / cost and loss of occupant comfort. We evaluate the time required to reach a stable control policy in addition to the question of robustness of control when faced with different environmental disturbances. The use of a simulator to generate building dynamics allows benchmarking the performance of the reinforcement learning agents against upper and lower bounds provided by MPC and RBC strategies respectively.

## Methodology

In order to test and benchmark the learning characteristics and abilities of the proposed RL algorithm, we used the building simulator described in (Ruelens F. , 2016). The simulator, implemented in Python, is an equivalent thermal parameter model (ETP), which simulates the heating and cooling of a building interior as a function of a limited number of lumped parameters, such as outdoor temperature and characteristics of the heating equipment. It is a deterministic second order model, which takes into account the heat being stored in the building envelope and causes it to cool down in a delayed manner in case of an outdoor temperature drop, or in the absence of introduced thermal energy. The building is considered equipped with a modulating air-source heat pump for space heating; a depiction of the control environment is in Fig. 1.

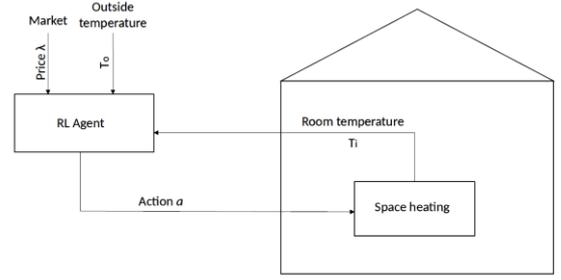

Fig. 1: Heat pump thermostat control environment (Ruelens F. , 2016)

**Influencing variables**

In the following, we list all the variables that influence the system dynamics as defined by the emulator:

- Ambient temperature $T_a^t$ is an exogenous variable and reflects the environmental disturbance in this scenario. It can't be influenced by the RL agent. For this research, five months of ambient temperature data was used and is visualized in Fig. 2.

- Building envelope temperature $T_{mass}^t$ is the latent energy embodied by the building at time instant t so it is not directly observable.

- Indoor temperature $T_i^t$ which is a function of the following:

$$T_i^t = f(T_a^{t-1}, T_{mass}^{t-1}, T_i^{t-1}, a^{t-1}) \qquad (1)$$

- The indoor temperature is measured and can be influenced by the control action chosen by the reinforcement agent, $a^{t-1}$.

- Control action $a^t$ reflects the input power of the heat pump at time $t$ which is a continuous value between 0 and $P_{max}$ [Watt].

- Energy consumption $c^t$ follows directly from the control action, $c^t = f(a^{t-1})[kWh]$.

- Energy price $\lambda^t$ is sampled from the price vector which is assumed to be known or, in case of real time pricing, a forecast is assumed to be available.

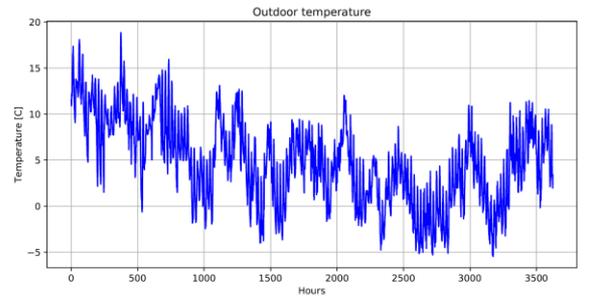

Fig. 2: Ambient temperature variation over 5 months

**Formulation as a Markov Decision Process (MDP)**

The reinforcement learning problem can be formulated as an MDP: $\{s, a, T, R\}$ (Sutton, 2017) where each individual element is defined as follows:

- State space: continuous, where $s^t \in S$ is the state at time t, consisting of n previous indoor temperatures including the current one where n is a parameter that can be optimized over and the current ambient temperature:
$$s^t: (T_i^t, T_i^{t-1}, T_i^{t-2}, \dots T_i^{t-n}, T_a^t) \quad (2)$$

- Action space: discretized between 0 and $P_{max}$, where $a^t \in A$:
$$a^t \in [0, 400, 800, 1200, 1600, 2000][W]$$

- Transition function, $T^t(s^t, a^t, s^{t+1})$ represents the system dynamics and is learnt from observation data

- Reward function $R^{t+1}(s^t, a^t, s^{t+1})$ represents the reward system that the agent has to maximize over time. The rewards are shaped so that higher priority is given to occupant comfort compared to energy or cost reduction. The negative signs reflect that this is a negative reward stream (or penalty). This reward stream is then split into these two components where the reward (or penalty) accrued because of energy consumption is given by:
$$R_{cons}^{t+1} = -c^t \lambda^t \quad (3)$$

Which depends on the price of energy consumption and the energy consumption itself. Loss of occupant comfort on the other hand is defined as:

$$R_{userComfortLoss}^{t+1} = \begin{cases} -3 \cdot 1.3^{T_i^t - T_{max}}, & \text{if } T_i^t > T_{max} \\ -4 \cdot 1.35^{T_{min} - T_i^t}, & \text{if } T_{min} > T_i^t \\ 0, & \text{otherwise} \end{cases} \quad (4)$$

Where $T_{max}$ and $T_{min}$ are the maximum and minimum comfortable temperature of the thermal zone. Fig. 3 shows the reward function. The exact numbers for the user loss reward function were derived empirically however a sensitivity analysis showed that the RL agent is capable of learning even if these rewards varied. The slightly asymmetric loss function derives from a review of thermal user comfort literature (De Dear, 2002), (ASHRAE, 2010). An additional reason for the asymmetry in the loss function derives from the fact that in our control problem the reinforcement learner has agency to heat the building and not cool it. Since we assume the building to be in a North-Western European climate, the loss of comfort in lower temperatures is much more relevant.

Based on ASHRAE studies, the optimal indoor temperature is 21℃ with a band of 2℃ which results in 90% occupant acceptability. Of course, the comfort range is extremely subjective and depends on individual preferences, something we take into account by making the reward system parametrized over choice of minimum and maximum acceptable temperatures. The RL agent is supposed to work for a range of user comfort valuations as long as it is prioritized over energy consumption. The total reward at any given time instant is then a simple summation of these two reward streams.

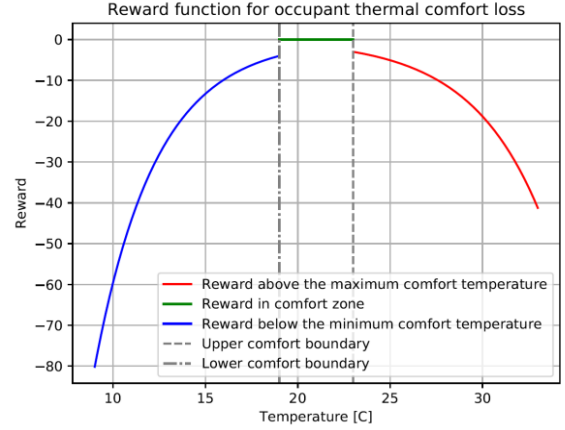

Fig. 3: Reward function for occupant comfort loss

The time horizon for optimization is 24 hours; the time step used by the emulator is one hour, which is also the resolution at which the agent issues new commands.

**Control objectives**

The reinforcement learning agent is expected to balance two reward streams. The first derives from respecting occupant comfort bounds while the second is to consume as little energy or money as possible while still meeting the first objective. The reinforcement agent has no prior information about the building dynamics at the beginning of control.

**Rule based control**

As mentioned earlier, the baseline controller implemented in many buildings today is rule based. This provides the lower bound on performance and serves as the benchmark on which the proposed algorithms have to improve over. This control takes the following form:
$$a_t = \begin{cases} P_{max} & \text{if } T_i^t < T_{min} - \Delta T \\ 0 & \text{otherwise} \end{cases} \quad (5)$$

Which implies that the heat pump starts consuming full power as soon as the temperature falls below the minimum threshold limit of occupant comfort by a predefined hysteresis band. There is no cooling functionality implemented so the agent takes no control action if the temperature exceeds the maximum allowable temperature no control action.

**Model predictive control**

A model predictive controller is implemented which assumes full knowledge of system dynamics. This provides an upper bound for the performance when compared with rule based control and the research question is how close a reinforcement learning based agent can approximate this solution.

**Model-based reinforcement learning**

The final benchmark we compare our model-free agent against is a model-based reinforcement learning agent. Here, the agent needs to learn the transition function to represent system dynamics from observed data. This is achieved using a neural network. This neural network takes as input the feature vector defined in eq. 2, and produces a scalar continuous valued output reflecting the indoor temperature at the next time step. It is possible to build future trajectories of arbitrary length by repeating this process multiple times. The model-based agent interleaves learning and planning to identify the action vector, which would maximize long-term reward. Multiple options were explored for the planning step. These include a cross entropy method (CEM) based planner (De Boer, 2005) and a genetic algorithm (GA) based planner (Fortin, 2012). Finally, the agent uses an $\varepsilon$-greedy algorithm to take exploratory steps to improve its learnt dynamics model. Here $\varepsilon$ is given by a harmonic sequence, which decays over time with $1/d^x$. The details of the implemented algorithm are summarized in pseudocode Algorithm 1.

**Algorithm 1** Model-based RL control with $\epsilon$-greedy exploration
1: Initialize sample memory $\mathcal{D}$ with capacity $N$, exploration rate $\epsilon$, transition ANN model $\widehat{T}_r$
2: **while** end of simulation **do**
3:    Initialize state $s$
4:    For next horizon $h$ obtain policy $\pi_h$ by *planning* using transition model $\widehat{T}_r$
5:    **for** k number of steps **do**
6:      With probability $\epsilon$ select random action $a$,
7:      otherwise select $a$ from $\pi_h$
8:      Execute $a$, observe $r$, $s'$, construct sample $S \leftarrow <s,a,s',r>$
9:      Store $S$ in $\mathcal{D}$
10:    **end for**
11:    With a daily frequency, improve transition ANN model $\widehat{T}_r$ on samples from $\mathcal{D}$ and update $\epsilon$
12: **end while**

The hyper-parameters of the neural network are chosen by grid search based on known design practices followed by tuning to obtain the best performance (Bengio, 2010), (F.-F. Li, n.d.).

**Model-free reinforcement learning (D-DNFQI)**

For the model-free algorithm, we use a variant of fitted Q iteration (Damien Ernst, 2005) which uses deep neural networks to approximate the optimal Q values for planning control actions:

$$Q(s,a,w) \approx Q^*(s,a) = Q^\pi(s,a) \qquad (6)$$

Here $w$ reflects the weight parametrizations of the neural network. The objective of this neural network is to minimize the mean squared error (MSE) with respect to the observed target value. Because of known issues with model-free learning such as convergence problems and instability in learning, target Q networks and prioritized experience replay are implemented in the algorithm (Volodymyr, 2013), (T. Schaul, 2015). Finally, to tackle the upward bias problem of Q estimations, Double Q learning is also implemented (V. Mnih, 2015). The final algorithm takes on the form of a double deep neural fitted Q iteration (D-DNFQI) with experience replay and $\varepsilon$-greedy exploration.

As before, the hyper-parameters of the neural network were optimized using search and existing guidelines. The structure of the output neural network was different than in the model-based RL case however. The Q-network directly output a 6 dimensional vector which reflected the 'goodness' of all possible control actions given an input state-action pair. The model-free algorithm is summarized in the pseudocode Algorithm 2.

**Algorithm 2** $\epsilon$-greedy Double Deep Neural Fitted Q Iteration (D-DNFQI)
1: Initialize experience replay memory $\mathcal{D}$ with capacity $N$, exploration rate $\epsilon$, network $Q$, target network $Q^-$,
2: **while** end of simulation **do**
3:    **while** end of horizon $h$ **do**
4:      Initialize state $s$
5:      With probability $\epsilon$ select random action $a$,
6:      otherwise $a \leftarrow \arg\max_a Q(s,a;w)$
7:      Execute $a$, observe $r$, $s'$, construct sample $S \leftarrow <s,a,s',r>$
8:      **if** $s'$ is not terminal state **then**
9:        Obtain target: $Tg(S) \leftarrow r + \gamma Q^-(s', \arg\max_{a'} Q(s',a';w); w^-)$ ▷ Double Q learning update rule
10:      **else**
11:        Obtain target: $Tg(S) \leftarrow r$
12:      **end if**
13:      Calculate $priority_S$ from $Tg(S)$ and $Q(s,a;w)$
14:      Store $S$ in $\mathcal{D}$ with $priority_S$
15:    **end while**
16:    *mini-batch* $\leftarrow \mathcal{D}$
17:    Train $Q$ network on *mini-batch*
18:    Update $w^-$ of $Q^-$ with $\tau$ and $w$
19:    Update each $priority_S$ in *mini-batch* using updated $Q$ and $Q^-$
20:    $\mathcal{D} \leftarrow$ updated *mini-batch*
21: **end while**

## Results

In this section, we benchmark the proposed algorithm to establish its efficacy against known controllers. For each test, the heating system is left uncontrolled with no input for 24 hours. After this initial period, the RL algorithms were given control whereupon learning started from scratch. In the following, we discuss performance of the proposed RL agent according to these pricing signals:

The case of **flat electricity pricing,** which translates to a purely energy efficiency concern.

The case of a **dual pricing scheme,** with fixed high prices during the day and fixed low prices during the night. This reflects the reality for many residential connections in countries with smart meters; the prices used were consumer tariffs from Belgium.

The case of **real time pricing,** where prices vary hourly on a daily basis. However, such pricing is usually valid only for aggregations of houses and not individual households, so we will not discuss it further in this paper.

We also consider boundary cases of interest after this discussion to evaluate the performance of different controllers on tasks where a classic model predictive controller would not be applicable. These include robustness to an incorrect model and to changing environmental constraints.

### Quality of model learnt in model-based learning

For model-based RL, the quality of the learnt model is of paramount importance. Since we are using a building simulation, we can directly compare against ground reality. To visualize the evolution of this model's performance over time, several snapshots were taken from the model at monotonically increasing times, i.e. as the neural network gathered more data. Fig. 4 shows that both the mean error and the variance around this error (obtained on an unseen test dataset) decays rapidly over time. The neural network exhibits acceptable performance after only a week's worth of experiences and after about 20 days the model has already learnt system dynamics almost perfectly.

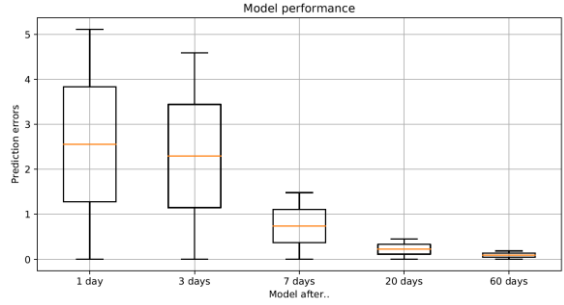

Fig. 4: Model performance over time (MAE) [°C]

**Flat price signal**

As discussed earlier, the results obtained have to be evaluated along two dimensions of interest: the impact on occupant comfort and the consequent energy or cost reductions.

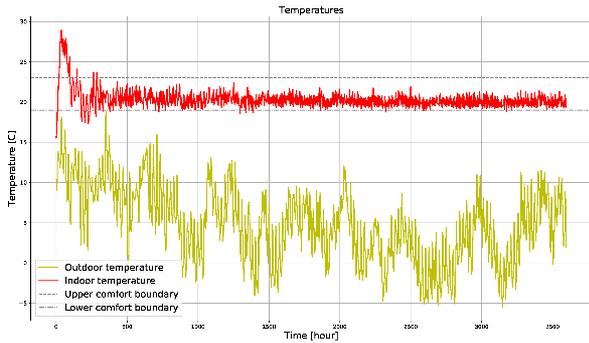

Fig. 5. Model-free RL performance in time, where the red and yellow plots show the indoor and ambient temperature respectively; acceptable band ranges from 19 - 23 degrees

Figs. 5 presents the indoor and outdoor temperature evolution for the period under consideration with the proposed model-free controller. It is obvious that the model-free controller exhibits a high peak indoor temperature during the exploration phase but then settles inside the comfort bounds quickly.

One difference that we observed between the model-free and the model-based controller is that the model-free controller tried to keep the indoor temperature as close to the lower comfort bound as possible, resulting in the heat pump providing some power to the building almost continuously. This is visualized in Fig. 6a and 6b where it can be seen that the heat pump is in the OFF state much less frequently for the model-free controller when compared with its model-based counterpart. While this increased costs, a useful side benefit arose in the sense that it consumes less peak power, which can be beneficial for the overall grid (assuming a number of similar heat pumps operating in a neighbourhood).

In addition to the very different control commands sent out by the reinforcement agent, it is also instructive to compare the two controllers with the lower and upper performance bounds obtained by rule based and ideal model predictive controllers respectively.

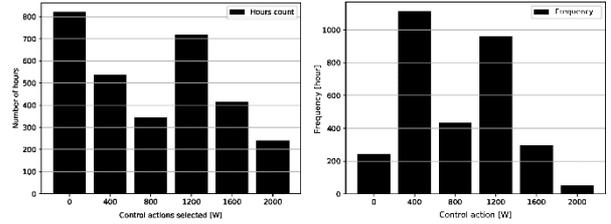

Fig. 6. Frequency of chosen control action by the RL agent for (a) Model-based; (b) Model-free

It is obvious from Table 1 that the model-free controller outperforms the rule-based controller in terms of cost reduction. At the same time, it is also evident that the model-free controller underperforms both the theoretical upper bound (achieved by perfect MPC) and the performance obtained by model-based RL. The loss of occupant comfort arose mostly during the initial training period where exploratory steps caused wild fluctuations in indoor temperature.

Table 1: Experimental results for a flat price profile

| Algorithm | Consumption change (%) | Cost change (%) | Loss of comfort (EUR) |
|---|---|---|---|
| **Rule based** | 0 | 0 | 0 |
| **Perfect MPC** | -8.8 | -8.8 | 0 |
| **Model-based RL** | -7.2 | -7.2 | 0.5 |
| **Model-free RL** | -6.4 | -6.4 | 5.23 |

**Dual price signal**

The learning problem becomes more challenging by switching from flat to dual tariffs. However, we observed similar behaviour (Fig. 7a and 7b). The model-free controller, as before, cycles between the ON states for the heat pump much more compared to the model-based one. The model-based controller, on the other hand, has learnt that keeping the heat pump OFF during the high price signal is desirable behaviour. The profiles on the histogram of the model-based controller make much more intuitive sense than the one for model-free control as the fraction of ON actions during low prices far outnumbers the fraction of ON actions for high price signal. However, as before the actions chosen by the model-free controller have a beneficial effect on the local low-voltage grid. This is not reflected in the costs shown in Table 2 however, which shows that, as before, while the model-free controller improves vastly on the rule-based controller; it is still not as efficient as the model-based controller.

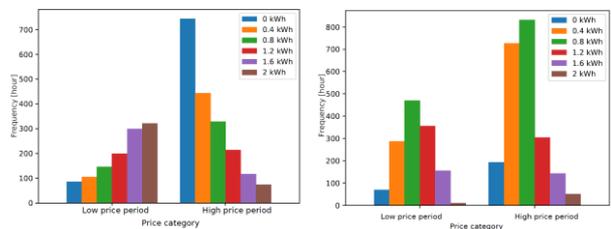

Fig. 7. Frequency of chosen control action by the RL agent for (a) Model-based; (b) Model-free

It is also interesting to note how consumption varies as a function of the profile in time. This is visualized in Fig. 8 where it is obvious that when the price is low, the heat pump is turned on more frequently. The two controllers have learnt strikingly different behaviour however. The model-based controller turns on the heat pump at (close to) full power as soon as the price shifts to low but seldom otherwise unless occupant comfort is at risk of being violated. The model-free controller offers a much more smooth response however with the majority of operation being in the mid-power regime regardless of the price point. There is a subtle shift however, with higher power control actions (1200W) prioritized when the price signal is low as compared to when it is high.

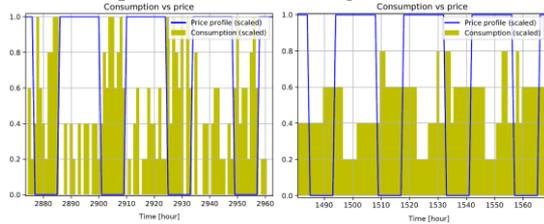

Fig. 8. Temporal behaviour of the RL agent with dual pricing for (a) Model-based; (b) Model-free

The table summarizing the reinforcement rewards as well as the cost and energy reductions can be seen below:

Table 2: Experimental results for a dual price profile

| Algorithm | Consumption change (%) | Cost change (%) | Loss of comfort (EUR) |
|---|---|---|---|
| Rule based | 0 | 0 | 0 |
| Model predictive | -5.0 | -10.7 | 0 |
| Model-based RL | -4.9 | -10.6 | 3.2 |
| Model-free RL | -7.7 | -8.0 | 8.0 |

**Robustness to changing constraints**

In real world settings, occupants frequently interact with the building thermal system to adjust the indoor climate according to their needs, by altering the temperature set point. So far, we have considered this set point as static. When occupants change this set point, the reinforcement learning agent has to adapt the policy it is following to make sure that it continues to perform in a desired manner. Fig. 9 illustrates the case for temperature set point that is first raised and then lowered before being set back to the original value. It is evident that model-based learning, where planning is decoupled from learning, quickly adapts to the new situation. Model-free learning however performs poorly because existing state-action pairs do not correspond to the updated Q-values anymore. By the time it has begun to learn the new representation, the constraints have changed again. This reflects an aspect of control where model-based learning is better suited than model-free control. However, given sufficient training data, the model-free controller can also learn optimal policies in this shifting constraints regime.

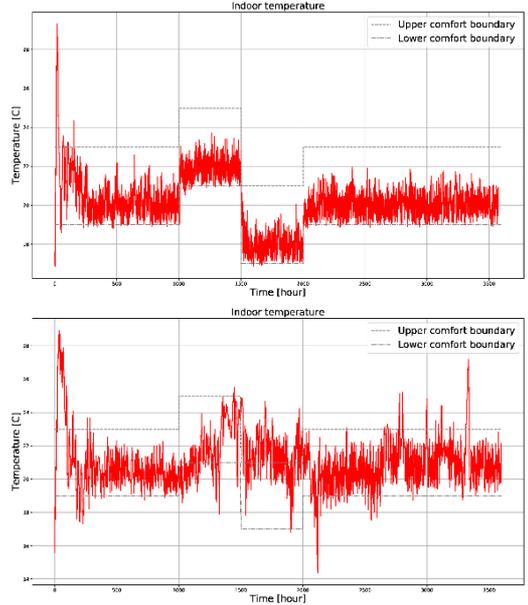

Fig. 9. Effect of temporally changing constraints for (a) Model-based; (b) Model-free RL

**Robustness to incorrect model**

In the previous section, we highlighted how the presence of a model helps the model-based controller to adapt quickly to dynamic conditions. This can, under some circumstances, also be its weakness. More specifically, problems can occur if the model learnt by the agent is incorrect or (as is more often the case) something changes in the environment.

To demonstrate this change in the simulation environment, the backup-controller of the heat pump has been turned on. This is a simple rule-based controller acting as a filter where control actions are overridden if the indoor temperature goes out of the comfort range, turning the heat pump on or off accordingly. $T_i$ is always kept around the desired limits, but the behaviour of the underlying filter also needs to be learned for simulating transitions. Results are shown in Fig. 10a and 10b for the model-based and model-free controllers. For the model-based agent, close to the comfort boundaries where actions are often overridden, the model fails to predict correctly and the search for optimal policy goes on a wrong trajectory. Since the model is never learned well during the simulation period, wrong commands are issued continuously, and the temperature is regulated entirely by the safety controller, resulting in continued comfort violations. As opposed to the model-based controller, model-free RL is less sensitive to changes of the environment dynamics. As seen in Fig. 10b, the controller had no problem learning a good policy and successfully steering the temperature within an acceptable temperature region.

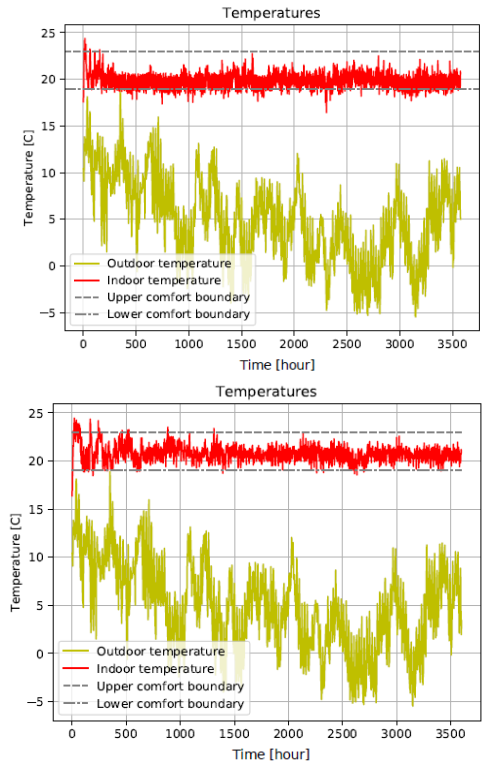

Fig. 10. Behaviour of the RL agent when the environment has changed for (a) model-based; (b) model-free

## Conclusions

This paper presents a novel model-free algorithm to perform space heating in a building. The system, as implemented here, is expected to be a good representation of reality; however, it does not consider complex occupant interactions. The objective of the controller was to steer the building climate in a way that preserved occupant comfort while reducing energy consumption and / or costs.

The results obtained with the proposed model-free controller were compared with a naïve but standard rule based controller and a model predictive controller, which assumes perfect information. These provide the upper and lower bounds for the performance achievable with reinforcement-based controllers. Based on the results in this study, it is obvious that such controllers can improve upon the performance of rule-based controllers and, in some cases, even approach the performance of perfect information MPC.

Most of the results obtained in this research point to the superiority of model-based controllers over model-free ones. This includes lower sample complexity (which translates to quicker learning) as well as higher reductions in cost and energy on average. Similarly, model-based reinforcement learning algorithms preserved occupant comfort much better in general and the system was resilient to changes in operating conditions.

This verifies conventional wisdom in control. However, the presented model-free algorithm does offer substantial benefits such as much faster compute times and the added benefit of working reliably when unexpected changes to environmental dynamics take place. There were also unexpected consequences as the model-free controller learnt a policy that was slightly costlier for the individual household, but generally better for grid stability.

To realize these cost and energy savings, we experimented with different price signals. Results vary with the choice of the price signal; however, the overall trends remain quite stable: costs can be reduced substantially with reinforcement learning strategies. These findings are summarized in Table 3.

Table 3: Comparison of key indicators of model-based and model-free RL

| Indicator | Model-based | Model-free |
|---|---|---|
| Sample complexity | 150-200 | 250-350 |
| Consumption reduction | 4.5-7% | 5-6% |
| Cost reduction | 7-18% | 5.5-10% |
| Exploration cost | + | ++ |
| Comfort loss | 0-7 EUR | 2-30 EUR |
| Computation time | 4-5 hours | 10-15 minutes |

These savings are possible because of the energy flexibility inherent in the thermal mass of buildings. By shifting consumption from times when prices are high to times when prices are low means that the building under consideration can be used for energy storage. The exact extent of this usage and its implications on using building thermal mass for providing services to the electric or thermal grid is an area for future consideration.

The issue of computational complexity is one avenue that might cause model-free algorithms to become more attractive for control. For real time planning, good control actions have to be generated in limited time horizons which might limit the applicability of model-based algorithms. Likewise, algorithms, which combine the computational cost of model-free and the accuracy of model-based algorithms, might lead to the next generation of building control. Similarly, the different strengths of these algorithms with regard to robustness means that a hybrid controller will most likely outperform either separately. This is another potential direction for future research.

Other avenues to extend the current research can proceed in multiple directions. The most promising is to replace the deterministic building emulator by a stochastic variant. Learning deterministic building dynamics through black-box models is possible because of the low dimensionality of the problem. This low dimensionality means that given enough time and interactions, the black box model will invariably learn the correct behaviour. A stochastic model on the other hand will more closely reflect real world situation where unobserved variables such as occupant behaviour produce unexpected changes in the building climate. To do this, different occupant profiles can be modelled and included as latent influencing variables in the simulation framework.

To conclude, this paper is an attempt at bringing more transparency to the data driven reinforcement learning algorithms that are increasingly found in literature as alternatives to established building control strategies. It is also likely that some variants of these algorithms will find a place in the multitude of smart thermostats and off-the-shelf building controllers. We have demonstrated here that while model-free algorithms underperform their model-based counterparts; they still offer substantial energy and cost savings when compared with rule-based controllers. That these savings come at only a fraction of the computational cost of model-based controllers is a heartening sign for future research.

## Acknowledgement

Hussain Kazmi gratefully acknowledges support and feedback from InnoEnergy, VLAIO and IEA Annex 67.